# Wood's anomalies and excitation of cyclic Sommerfeld resonances under plane wave scattering from a single dielectric cylinder at oblique incidence of light

Andrey D. Pryamikov[1] and Alexander S. Biriukov

[1]Fiber Optics Research Center of Russian Academy of Sciences, 38 Vavilov street, Moscow, 119333, Russia

[*]Corresponding author: pryamikov@fo.gpi.ru

In this paper we consider a process of plane wave scattering from a single dielectric cylinder at oblique incidence under which it is possible to excite long range cyclic Sommerfeld waves (CSWs) and consequently cyclic Sommerfeld resonances (CSRs) of different orders. It is shown that the CSRs are analogous to Wood's anomalies occurred under plane wave scattering from one dimensional (1D) metallic diffraction gratings. The conditions which are necessary for an effective excitation of CSWs and CSRs with high quality factors are analyzed. 2012 Optical Society of America

## Introduction

The plane wave scattering from a single dielectric cylinder of infinite length at oblique incidence has been considered by many authors. For the first time, a solution of the problem was given in [1]. The amplitudes of T. M. and T. E. modes of scattered and internal fields were determined and some special cases of the scattering were considered. Later, in [2] these solutions were applied for determining radiation modes of the single dielectric cylinder and were called I. T. M. (incident transverse magnetic) and I. T. E. (incident transverse electric) continuous modes. The authors of [3] calculated numerically scattering coefficients and extinction coefficients for obliquely oriented infinite circular cylinder. They have shown that the extinction coefficients for oblique incidence have much more intense resonance structure than those for normal incidence. In our work we will consider special cases of the plane wave scattering from a single dielectric cylinder at oblique incidence of light under which it is possible to excite CSWs and CSRs of different orders. It is known [1] that components of the scattering and the internal fields as well as the incident one can be expanded into an infinite series according to the addition theorem for Bessel functions [4]. All terms of the series have an azimuthal dependence $e^{\pm in\varphi}$, where $\varphi$ is an azimuth in a cylindrical coordinate system and $n$ is an integer. It is generally assumed that the corresponding harmonics amplitudes of the same order $a_n$ and $a_{-n}$ are not connected with each other by any additional relation. In such a way, they can be represented as two independent counter propagating cylindrical harmonics (waves) with the same propagation constant. We will show that special additional relations between these amplitudes give rise to an excitation of the CSWs on the cylinder surface. For the first time, an excitation of the CSWs and corresponding CSRs occurred under plane wave scattering from a dielectric and metallic cylinders were discussed in [5]. The authors considered a plane wave scattering at grazing angles from the dielectric and metallic nanowires. They showed that it is possible to excite the long range guided surface waves (CSWs) through interplay of coherently scattered continuum modes which couple with first order azimuthal propagating modes of the nanowires. As it will be shown further, there is a necessary condition for such coupling which is that to achieve a phase – matching condition between continuum modes and the propagating mode not only in the propagation direction but also in the azimuthal direction. The CSRs occurred under such excitation have a much higher $Q$ factor than the pure plasmonic resonances on the surface of metallic nanowires.

In this paper we will analyze an excitation of CSWs and resonance structures of CSRs occurred under plane wave scattering from a single dielectric cylinder at oblique incidence and show that this phenomenon has a close association with resonances occurred under plane wave scattering from 1D metallic diffraction gratings. Such resonances are known as Wood's anomalies [6]. For the first time, this phenomenon was observed by Prof. Wood in 1902. He discovered the presence of unexpected narrow bright and dark bands in the spectrum of a metallic diffraction grating illuminated by light source with a slowly varying spectral intensity distribution. It was found that the occurrence of these bands was dependent on polarization of incident light and the groove depth of the grating. The first theoretical explanation of the anomalies was given by Lord Rayleigh in 1907 [7, 8]. Using Huygens' construction he showed [7] that each groove of the grating is an origin of the scattered spherical wave. Rayleigh's condition is that there exist diffracted waves at grazing emergence. It means that the part of the scattered wave travels along the grating and reaches the neighboring groove in phase with the incident wave and with the waves scattered by other grooves. As a result, the second order diffraction of the scattered waves interferes with the first order diffraction of the incident wave to yield Wood's anomalies. In work [8] the scattered electromagnetic fields were expanded in terms of the outgoing waves only. In this interpretation the scattered fields became singular at wavelength for which one of the spectral orders emerges from the grating at grazing angles. The wavelengths at which these singularities occur are called Rayleigh' wavelengths $\lambda_R$ and correspond to the Wood's anomalies. One of the main disadvantages of Rayleigh's theory was that it was impossible to obtain the shape of the bands associated with the anomalies. U. Fano in [9] introduced a method of successive approximations to explain the anomalous interference effects and gave an explanation of the anomalies in terms of superficial waves excited on the metallic grating surface. These superficial waves can be considered as quasi stationary superficial waves obtained mathematically by Sommerfeld (Sommerfeld's waves). In connection with this problem U. Fano in [9] told that the problem of propagation of light traveling along 1D diffraction grating is analogous to the problem of the radiation by an antenna near the surface of a conducting body. The similar point of view on the origin of Wood's anomalies was proposed in [10]. Contrary to customary multiple scattering procedures [11] the authors of [10] presented a theory of Wood's anomalies which is based on guided wave approach. They showed that there are two types of the anomalies: a Rayleigh type of the ones which is characterized by an abrupt intensity modulation of the diffraction orders at appearance and fading of the new spectral orders and a resonant type of the anomalies corresponding to the resonances of guided complex waves of the gratings. It is worth mentioning that these two types of the anomalies can occur either separately or even simultaneously. In the next sections we will compare resonance phenomena occurred under plane wave scattering from a single dielectric cylinder at oblique incidence and from 1D diffraction grating based on this work [10]. For the first time, Fano like resonances occurred under plane wave scattering from the single dielectric cylinder at oblique incidence were described in [12]. The authors of [12] have shown that Fano like resonances excited in such a way appear in the loss spectra of all solid photonic band gap fibers. In their opinion, such behavior of the loss spectrum is possible when ratio of $d/\Lambda$ is small, where $d$ is a diameter of the cylinder and $\Lambda$ is a pitch, and the coupling between cylinders of the cladding is weak. In this case, the scattering properties of the individual cylinder determine the spectral properties of the all solid photonic band gap fiber according to the ARROW model [13, 14].

In the end of this section we will discuss some modern aspects of Wood's anomalies theory. Ebbesen and coworkers [15] explored optical properties of arrays of submicrometer cylindrical cavities in metallic films and reported an unusual zero – order transmission spectra at wavelengths larger than the array period. They observed the enhanced transmission through the metallic film at wavelengths as large as ten times the diameter of the cylindrical cavities. The authors of [16] provided experimental evidence that the enhanced transmittance of light through a silver film pierced by a periodic array of subwavelength holes assisted by the enhanced fields associated with surface plasmon polaritons. In work [17] the authors have shown numerically that transmission and reflection behavior of light under the optical transmission on arrays of subwavelength holes in a metallic layer deposited on a dielectric substrate correspond to Fano's profile. The authors also showed that such form of the profile occurs due to the resonant response of the egeinmodes of the structure coupled with nonhomogeneous diffraction orders and the transmission properties of the structure result from the resonant Wood's anomalies. Beginning with the first works [15, 16] on the enhanced optical transmission through thin metallic films containing periodic arrays of subwavelength holes it was assumed that surface plasmons should play a key role in the observed transmission spectrum features. Later it was found that the origin of the eigenmodes is not required to account for the general features of the transmission [17]. For example, in [18] the authors discussed and demonstrated numerically the enhanced transmission through an array of subwavelength cylindrical holes in a tungsten layer. The tungsten exhibits a positive permittivity real part in the considered wavelength region and surface plasmons cannot exist but the transmission features were found very similar to that obtained with the metallic films. An example of the enhanced transmission through a chromium film in the wavelength region where the dielectric constant is positive was demonstrated in [19]. The authors have shown numerically that the eigenmodes involved in such transmission are not surface plasmons no guided modes. They were substituted by Brewster – Zennek modes. In other words, this behavior also correlated with resonant response of the eigenmodes coupled with nonhomogeneous diffraction orders and can also be explained as resulting from the resonant Wood's anomalies. In such a way, this problem is very closely connected to the problem of Wood's anomalies occurred under the plane wave scattering from 1D metallic diffraction gratings. It is worth mentioning that the authors of [20] using a theoretical formalism in both real and $k$ spaces have shown that the physical origin of the extraordinary transmission of light through quasiperiodic arrays of holes connects to an existence of long range orders in the quasiperiodic arrays. These long range orders select the wavevectors of the surface electromagnetic modes which allow the extraordinary transmission through subwavelength holes.

Also, it is necessary to point out that there are two types of resonances for such systems, namely, the resonances connected with an internal property of the single scatterer and multiple scattering dynamical diffraction geometrical resonances [21]. The first type of the resonances is originated from an existence of the particle surface modes or plasmons which induce peaks in the reflectance. The second type of the resonances is determined by the optical properties of an array of the scatterers. Both types of the resonances can be called as Fano – Feshbach resonances and they occur when the frequency of the incoming wave is tuned to the frequency of a quasi bound states described above [21]. In this paper only the first type of the resonances will be considered.

As it was mentioned above, CSWs and, consequently, CSRs occurred under the plane wave scattering from a single dielectric cylinder are due to the coupling between coherently continuum modes of the single cylinder and its first order azimuthal propagating eigenstates. It can be stated that CSWs and CSRs are formed by analogy with the phenomena of extraordinary transmission of light through the arrays of subwavelength holes and based on the same physical principles. In such a way, the origin of CSWs occurred under plane wave scattering from a single dielectric cylinder and the shapes of CSRs can correlate with the origin and shapes of Wood's anomalies occurred in such different physical systems as 1D metallic diffraction grating or the arrays of subwavelngth holes. This work is devoted to understanding a deep connection between these phenomena, namely, the origin of the resonances occurred under plane wave scattering from a single dielectric cylinder and from 1D metallic diffraction gratings.

The paper is organized as follows. In Section 2, we consider the problem of plane wave scattering from a single dielectric cylinder and CSRs excitation. Then, we compare these results with the ones obtained in [10] for 1D metallic diffraction gratings and draw an analogy between these two phenomena. In section 3, we carry out numerical analyses of the resonance structures of CSRs and compare them with the resonances with Lorentzian and Fano profiles described in [10]. In Section 4, we analyze the conditions which are necessary for an effective excitation of CRSs and CSWs. Section 5 contains the conclusions.

**An excitation of cyclic Sommerfeld waves on the surface of dielectric cylinder and plane wave scattering from one dimensional diffraction grating**

A solution of the problem of plane wave scattering from a single dielectric cylinder is well known and we will briefly discuss the main points. Let us consider a TE polarized plane wave incident on the cylinder with radius of $a$ and with refractive index of $n_1$. The case of TM polarization can be considered analogously. The refractive index of the outer region is $n_2$. In this case, $z$ – components of magnetic and electric fields is given by:

$$H_z^i = H_0 \sin\theta e^{ik_2 \sin\theta \cos\varphi \rho} e^{-ik_2 \cos\theta z} e^{i\omega t}, \tag{1}$$

$$E_z^i = 0,$$

on the assumption that the magnetic vector is parallel to the plane $\varphi = 0$, where $\varphi$ is an azimuth in cylindrical polar coordinate system ($\rho$, $\varphi$, $z$). In (1) $\theta$ is an angle of incidence, $k_2 = 2\pi n_2 / \lambda$ is a wave vector of the incident plane wave (momentum of the wave) and $\lambda$ is a wavelength of the incident light. Employing the addition theorem for Bessel functions the component of the magnetic field can be written [1, 4]:

$$H_z^i = H_0 \sin\theta \sum_{n=-\infty}^{+\infty} i^n J_n(q_2 \rho) e^{-in\varphi} e^{-i\beta z} e^{i\omega t}, \tag{2}$$

where $q_2 = \sqrt{k_2^2 - \beta^2}$ and $\beta = k_2 \cos\theta$ is a propagation constant.

In the following we will omit for convenience the term in (2) connected with propagation in $z$ – direction $e^{-i\beta z}$ and the time dependence term of $e^{i\omega t}$. Other tangential components of the incident electric and magnetic fields can be deduced from $z$ – components of the ones by using well known relations [22]:

$$H_\varphi^i = -\frac{i}{q_i^2} \left( \frac{\beta}{\rho} \frac{\partial H_z^i}{\partial \varphi} + \omega \varepsilon_0 n_i^2 \frac{\partial E_z^i}{\partial \rho} \right) \tag{3}$$

$$E_\varphi^i = -\frac{i}{q_i^2} \left( \frac{\beta}{\rho} \frac{\partial E_z^i}{\partial \varphi} - \omega \mu_0 \frac{\partial H_z^i}{\partial \rho} \right),$$

where $\varepsilon_0$ is a permittivity of free space, $\mu_0$ is a permeability of free space and $n_i$ is a refractive index ($i = 1,2$). The scattered and the internal fields have $z$ – components which can be presented in the same form as in (2):

$$H_z^s = \sum_{n=-\infty}^{+\infty} a_n^s H_n^{(2)}(q_2\rho)e^{-in\varphi} \qquad (4)$$

$$E_z^s = \sum_{n=-\infty}^{+\infty} b_n^s H_n^{(2)}(q_2\rho)e^{-in\varphi}$$

$$H_z = \sum_{n=-\infty}^{+\infty} a_n J_n(q_1\rho)e^{-in\varphi}$$

$$E_z = \sum_{n=-\infty}^{+\infty} b_n J_n(q_1\rho)e^{-in\varphi} ,$$

where $q_1 = \sqrt{k_1^2 - \beta^2}$ and $k_1 = 2\pi n_1/\lambda$. The $\varphi$ - components of the scattered and the internal fields can be found in an analogous way as in the case of the incident fields (3).

As it was shown in [2] the continuous mode spectrum of a circular dielectric rod can be constructed by using the fields having a structure as in (2) – (4). These modes can be I. T. E. type when $E_z^i = 0$ or I. T. M. type when $H_z^i = 0$ as it was mentioned above. In other way, these modes can be called radiation modes of the cylinder [23]. There is no any additional connection between the coefficients $a_n^s$ and $a_{-n}^s$ or $b_n^s$ and $b_{-n}^s$ for the continuous (radiation) modes [2]. Thus, two azimuthally counter – propagating helical harmonics (waves) with opposite signs in the azimuthal dependencies of $e^{in\varphi}$ and $e^{-in\varphi}$ travel along the cylinder independently from each other. Analogous situation is observed for the scattered fields in the case of 1D metallic diffraction gratings when two diffraction orders propagate with the dependencies of $e^{-i\beta(z+\frac{2\pi m}{d})}$ and $e^{-i\beta(z-\frac{2\pi i}{d})}$, where $d$ is a period of the diffraction grating. On the other hand, the expression (2) for $z$ – component of magnetic field of the incident wave can be rewritten in the form:

$$H_z^i = H_0 \sin\theta \left( J_0(q_2\rho) + 2\sum_{n=1}^{\infty} i^n J_n(q_2\rho)\cos(n\varphi) \right) \qquad (5)$$

$$E_z^i = 0,$$

from which it is seen that $z$ – component of the magnetic field can be represented as a standing wave in azimuthal direction. Such expansion is possible just due to the properties of Bessel functions $J_{-n}(\lambda_2\rho) = (-1)^n J_n(\lambda_2\rho)$. According to (5) all other $z$ – components of the scattered and the internal fields (4) can be represented in the same form as in (5):

$$H_z^s = a_0^s H_n^{(2)}(q_2\rho) + 2\sum_{n=1}^{+\infty} a_n^s H_n^{(2)}(q_2\rho)\cos(n\varphi) \qquad (6)$$

$$E_z^s = -2i\sum_{n=1}^{+\infty} b_n^s H_n^{(2)}(q_2\rho)\sin(n\varphi)$$

$$H_z = a_0 J_0(q_1\rho) + 2\sum_{n=1}^{+\infty} a_n J_n(q_1\rho)\cos(n\varphi)$$

$$E_z = -2i\sum_{n=1}^{+\infty} b_n J_n(q_1\rho)\sin(n\varphi).$$

Such representation is possible on the assumption that the coefficients of the scattered fields and the internal fields (4) are connected between each other in the next form:

$$a_n^s = (-1)^n a_{-n}^s \qquad (7)$$

$$b_n^s = (-1)^{n+1} b_{-n}^s,$$

$$a_n = (-1)^n a_{-n},$$

$$b_n = (-1)^{n+1} b_{-n}.$$

As it was shown in [5] the harmonics (waves) traveling along the cylinder with $z$ – components representing in the form of (5) – (6) at near – zero grazing incidence can be considered as long range guided surface waves. They are originated from coupling between the continuous (radiation) modes of the dielectric cylinder and the first order azimuthal propagating modes of the dielectric cylinder. Such waves were called cyclic Sommerfeld waves and cyclic Sommerfeld resonances are originated from the excitation of these waves [5].

The expansion coefficients in (5) – (6) are obtained by solving corresponding systems of four linear equations for TE polarized incident wave:

$$\begin{pmatrix} H_n^{(2)}(q_2 a) & 0 & -J_n(q_1 a) & 0 \\ 0 & H_n^{(2)}(q_2 a) & 0 & -J_n(q_1 a) \\ \dfrac{i\beta n}{q_2^2 a} H_n^{(2)}(q_2 a) & -\dfrac{\omega n_2^2 \varepsilon_0}{q_2} H_n^{(2)'}(q_2 a) & -\dfrac{i\beta n}{q_1^2 a} J_n(q_1 a) & \dfrac{\omega n_1^2 \varepsilon_0}{q_1} J_n'(q_1 a) \\ \dfrac{i\omega\mu_0}{q_2} H_n^{(2)'}(q_2 a) & -\dfrac{\beta n}{q_2^2 a} H_n^{(2)}(q_2 a) & -\dfrac{i\omega\mu_0}{q_1} J_n'(q_1 a) & \dfrac{\beta n}{q_1^2 a} J_n(q_1 a) \end{pmatrix} \begin{pmatrix} a_n^s \\ b_n^s \\ a_n \\ b_n \end{pmatrix} = \begin{pmatrix} -H_0 \sin\theta i^n J_n(q_2 a) \\ 0 \\ -H_0 \sin\theta \dfrac{i^{n+1}\beta n}{q_2^2 a} J_n(q_2 a) \\ -H_0 \sin\theta \dfrac{i^{n+1}\omega\mu_0}{q_2} J_n'(q_2 a) \end{pmatrix} \quad (8)$$

This system is obtained from the boundary conditions for $\varphi$ and $z$ – components of the fields. It is seen that the amplitudes of the magnetic and electric fields of $n$th harmonic in (8) will be dependent on $n+1$ and $n-1$ harmonics due to a property of derivatives of Bessel functions of the first kind and the Hankel functions [4]. Solving (8) for the scattered field coefficients one can obtain:

$$a_n^s = H_0 \sin\theta i^n \frac{J_n(y)}{H_n^{(2)}(y)} \frac{\left\{\left(\dfrac{1}{y}\dfrac{J_n'(y)}{J_n(y)} - \dfrac{1}{x}\dfrac{J_n'(x)}{J_n(x)}\right)\left(\dfrac{n_1^2}{x}\dfrac{J_n'(x)}{J_n(x)} - \dfrac{n_2^2}{y}\dfrac{H_n^{(2)'}(y)}{H_n^{(2)}(y)}\right) + \dfrac{\beta^2 n^2}{k_0^2}\left(\dfrac{1}{x^2} - \dfrac{1}{y^2}\right)^2\right\}}{\left\{\left(\dfrac{1}{x}\dfrac{J_n'(x)}{J_n(x)} - \dfrac{1}{y}\dfrac{H_n^{(2)'}(y)}{H_n^{(2)}(y)}\right)\left(\dfrac{n_1^2}{x}\dfrac{J_n'(x)}{J_n(x)} - \dfrac{n_2^2}{y}\dfrac{H_n^{(2)'}(y)}{H_n^{(2)}(y)}\right) - \dfrac{\beta^2 n^2}{k_0^2}\left(\dfrac{1}{x^2} - \dfrac{1}{y^2}\right)^2\right\}}, \quad (9a)$$

$$b_n^s = \frac{\left(H_0 \sin\theta i^n \dfrac{J_n(y)}{H_n^{(2)}(y)} + a_n^s\right)\left(\dfrac{1}{q_1^2} - \dfrac{1}{q_2^2}\right)}{\left(\dfrac{n_1^2}{x}\dfrac{J_n'(x)}{J_n(x)} - \dfrac{n_2^2}{y}\dfrac{H_n^{(2)'}(y)}{H_n^{(2)}(y)}\right)}\left(\dfrac{i\beta n}{\omega\varepsilon_0}\right), \quad (9b)$$

where $x = q_1 a$ and $y = q_2 a$. It will be shown that the resonances of the coefficients will be determined as by the poles of $1/J_n(x)$ as by more complicated combinations of the functions $J_n(x)$, $J_{n+1}(x)$ and $J_{n-1}(x)$.

It is appropriate at this point to draw again the analogy between the plane wave scattering from a single dielectric cylinder and from the 1D metallic diffraction grating. For the first time, we have drawn such analogy in [24] when transmission of IR radiation through hollow core microstructured fibers was considered. Here, we will consider the problem in more detail. Let us consider briefly a scattering of light from 1D diffraction grating based on the approach described in [10]. If the periodically modulated reactance plane extends infinitely in $z$ and $y$ direction, with period $d$ in $z$ direction only, the surface impedance $Z^s(z)$ can be represented by the Fourier series expansion:

$$Z^s(z) = \sum_{n=-\infty}^{+\infty} Z_n^s e^{i(2\pi n/d)z} \quad . \tag{10}$$

Suppose the plane wave with its magnetic field parallel to the $y$ axis is incident on the plane at angle $\theta$ with respect to the normal to the plane:

$$H_y^i = H_0 e^{iqx} e^{-i\beta z} \quad ,$$

where $\beta = k\cos\theta$, $q = \sqrt{k^2 - \beta^2}$ and $k = \omega/c$ is the free space wavenumber. Due to periodicity, the scattered magnetic field can be represented in the form of a Floquet type expansion:

$$H_y^s(x,z) = \sum_{n=-\infty}^{+\infty} A_n(\beta) e^{iq_n x} e^{-i2\pi n z/d} e^{-i\beta z} \quad , \tag{11}$$

for $x \geq 0$ and $n = 0, \pm 1, \ldots$ and $q_n = \sqrt{k^2 - (\beta + 2\pi n/d)^2}$. Using the periodic impedance boundary conditions at the grating plane $z = 0$ one can obtain an infinite set of simultaneous linear equations for the scattered field amplitudes [10]:

$$\sum_{m=-\infty}^{+\infty}\left(Z_{n-m}^s + q_n/(\omega\varepsilon)\delta_{nm}\right)A_m = 2q/(\omega\varepsilon)H_0\delta_{n0} \quad , \tag{12}$$

where $\delta_{n0}$ is Kronecker symbol. Then, the set of equations (12) can be expressed in the compact form as:

$$(Z)A = V$$

and the amplitudes $A_n$ are:

$$A_n = \frac{\Delta_n}{\Delta}, \tag{13}$$

where $\Delta$ is the determinant of the matrix $(Z)$ and $\Delta_n$ is the formal infinite determinant which is obtained by replacing $n$th column in $\Delta$ by the column vector $V$. According to [9, 10] there are two anomalous effects occurred under plane wave scattering from the 1D metallic diffraction grating. The resonance anomaly is connected with solutions of the homogeneous equations and consequently with the vanishing of determinant $\Delta$ in (13). It is known that such type of anomaly corresponds to an excitation of leaky waves of the grating with complex propagation constant $\beta^s = \beta_{Re}^s + i\beta_{Im}^s$ and having Lorentzian profile with a finite value of $Q$ factor ($Q = \beta_{Re}^s/2\beta_{Im}^s$). If $n$th spectral order of the scattered wave with an amplitude $A_n(\beta)$ has a propagation constant $\beta + 2\pi n/d \approx \beta_{Re}^s$ this spectral order couples to the leaky wave of the grating and give rise to the resonance anomaly in the considered spectral range. The second type of the anomalies is connected with a rapid variation in the amplitude of the diffracted spectral orders, corresponding to the onset or disappearance of a particular spectral order. For example, it can be demonstrated by considering $(n-1)$ spectral order with an amplitude $A_{n-1}(\beta)$ if the determinant $\Delta_{n-1}$ in (13) posses a zero near to the value of $\beta + 2\pi n/d$ at which the expression (13) has a complex pole. In this case, the expression (13) contains the product of terms involving a simple complex pole and a first order complex zero near to each other but not at identical location. It means that the expression (13) is a function which has a maximum and a minimum in its proximity. Such resonance anomalies are associated with Rayleigh wavelengths.

In [9] U. Fano told that the whole set of diffraction waves may be determined representing diffraction as a momentum transfer by 1D diffraction grating to the impinging wave. In the case of diffraction grating this momentum is transferred to the impinging wave in *z* or opposite direction. According to U. Fano, the pairs of waves with tangential momentum components $k_{pt} = k_0 \pm 2\pi m/\Lambda$ excited by the incident waves on the surface of 1D diffraction grating can be considered as forced oscillations. These oscillations can oscillate intensively as quasi stationary wave if the tangential momentum component of $k_{pt}$ approaches the one of the quasi stationary wave. Moreover, this problem is very similar with a problem of the radiation by an antenna near the surface of a conducting body considered by Sommerfeld. As in the case of Sommerfeld's problem, the spherical wave diffracted by a groove is analogous to the wave irradiated by a resonating oscillator lying just above the groove [9]. According to Fano's explanation the waves scattered along the surface of the grating act as a single damped wave with tangential momentum component corresponding to the superficial quasi stationary wave. On the basis of this statement, Fano made a conclusion that in the case of 1D diffraction grating the quasi stationary wave (Sommerfeld's wave) can be represented as a group of undamped waves whose interference appears as a damping [9].

At the same time, the Fourier series expansions (2) and (4) may be determined representing diffraction as a momentum transfer by the cylindrical surface to the incident plane wave if one considers the cylindrical surface as quasi diffraction grating with a period $2\pi$ in $\varphi$ - direction (azimuth). In this case, the scattering of TE polarized plane wave incident on the cylinder surface corresponds to the scattering of TM polarized plane wave incident on 1D diffraction grating considered above. One can state that the scattering from these two different physical systems is to transfer a momentum to the impinging plane wave in two different ways and in two different directions. One way of the momentum transfer to the impinging plane sw ave is connected with a periodicity of 1D diffraction grating into the wave propagation direction and with corresponding Fourier expansions of the electromagnetic fields. Another way of the momentum transfer is connected with the cyclic periodicity of cylindrical surface into azimuthal direction which allows to obtain analogous Fourier expansions for electromagnetic fields in the cylindrical coordinates. Two helical waves with the azimuthal dependencies of $e^{in\varphi}$ and $e^{-in\varphi}$ can interfere with each other to give rise the quasi stationary cyclic Sommerfeld wave in the same way as in the case of 1D metallic diffraction grating. In such a way, the resonance effects for the scattered fields occurred under such momentum transfer should be analogous to Wood's anomalies occurred under plane wave scattering from 1D metallic diffraction grating.

The main difference between these two types of the momentum transfer is that the expansion (2) can be represented into the form of (5) (as a standing wave $\cos(n\varphi)$ or $\sin(n\varphi)$ in azimuthal direction) due to the properties of cylindrical functions $J_n(x)$ and $H_n^{(2)}(y)$. To excite Sommerfeld wave in the case of 1D metallic diffraction grating it is necessary to fulfill phase matching condition between the propagation constant of $n$th scattered harmonic (wave) $\beta + 2\pi n/d$ and propagation constant of the leaky wave of the grating $\beta_{\text{Re}}^s$. To excite CSW which occurs due to interplay between the scattered continuum modes of $n$th order and propagating modes of the cylinder of the same order it is necessary to fulfill phase matching condition not only in $z$ – direction as in the case of 1D diffraction grating but also in $\varphi$ - direction in cylindrical coordinates. In such a way, the formation of the standing wave in azimuthal direction can be understood in terms of phase matching condition in $\varphi$ - direction for the scattered continuum modes with azimuthal dependencies of $e^{in\varphi}$, $e^{-in\varphi}$ and propagating mode of the dielectric cylinder which has the azimuthal dependencies $\cos(n\varphi)$ or $\sin(n\varphi)$ [23]. The excitation of CSW occurs at the resonance wavelengths when both phase matching conditions are fulfilled.

The second difference between these two types of momentum transfer is that the scattered waves and the incident wave in the case of 1D diffraction grating have the same type of polarization while in the case of plane wave scattering from the single dielectric cylinder at oblique incidence the scattered waves have nonzero $z$ - components of electric and magnetic fields [1]. It is occurred because the momentum is transferred from the cylinder surface to the incident wave into $\varphi$ - direction and $z$ – components of harmonics of the scattered electric field (6) are not equal to zero under incidence of TE polarized plane wave (1), (4).

In the next section we will study the profiles of CSRs occurred under CSWs excitation and show that they correspond to the Wood anomalies occurred under the plane wave scattering from 1D metallic diffraction grating.

## Cyclic Sommerfeld resonances and Wood anomalies

Let us consider the scattering of the plane wave at oblique incidence from a single dielectric cylinder with a refractive index $n_1 = 1.478$ and with a radius of $a = 1.2$ μm. The refractive index of the outer medium is $n_1 = 1.45$. It is assumed that CSWs can be excited on the cylinder surface and the conditions (7) must be held. One calculates spectral dependencies of the absolute values of coefficients $a_n^s(\lambda)$ of the scattered magnetic field (9a). In Fig.1 the dependencies for $n = 0, 1, 2, 3, 4$ are shown at an angle of incidence of the plane wave $\theta = 0.86^0$.

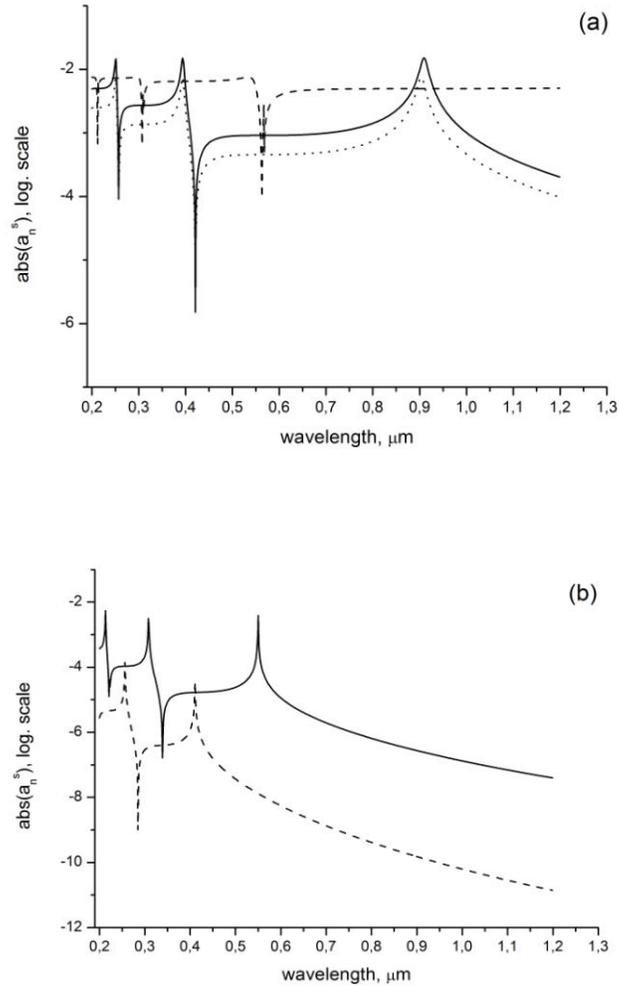

Fig. 1. (a) the spectral dependencies of $a_n^s$ (9a), $n = 0$ (solid), $n = 1$ (dashed), $n = 2$ (dotted); (b) $n = 3$ (solid), $n = 4$ (dashed).

In the case of TM polarized incident wave the structures of the resonances is the same and we don't consider them here.

As one can see from Fig. 1, the structure of the resonances of the excited CSWs are rather complicated and are very similar to the ones obtained in [10] for the Wood anomalies excited under scattering of plane wave from 1D metallic diffraction grating. In Fig. 1 there are resonances as with Lorentzian profiles as with Fano like profiles. To explain such behavior of CSRs it is necessary to consider the expressions (9(a)) for $a_n^s$ more thoroughly. First, one considers the spectral dependence of $abs(a_0^s)$ which has a form:

$$a_0^s = H_0 \sin\theta \frac{J_0(y)}{H_0^{(2)}(y)} \frac{\left(\frac{1}{y}\frac{J_0'(y)}{J_0(y)} - \frac{1}{x}\frac{J_0'(x)}{J_0(x)}\right)}{\left(\frac{1}{x}\frac{J_0'(x)}{J_0(x)} - \frac{1}{y}\frac{H_0^{(2)'}(y)}{H_0^{(2)}(y)}\right)} \quad . \tag{14}$$

It is seen from Fig. 1(a) that the spectral dependence of $abs(a_0^s)$ has CSR with Lorentzian profile in the long wavelength spectral region and CSRs with Fano like profiles in the shorter wavelength regions. To understand the origin of these CSRs profiles it is necessary to consider the absolute values of the denominator and the numerator in (14) separately. In Fig. 2 the spectral dependencies of the denominator and the numerator are shown for comparison with the spectral dependence of $abs(a_0^s)$. The locations of the resonances in the spectral dependence of the numerator (14) coincide with the ones in the spectral dependence of $abs(a_0^s)$. All CRSs have Fano like profile apart from the one located at λ = 0.9 μm which has Lorentzian profile. At the same time, all CSRs in the spectral dependence of the denominator have Fano like profiles (Fig. 2). To explain such behavior of the CSRs it is necessary to consider interference between the terms in the denominator and numerator in (14) corresponding to the incident fields and the fields inside the cylinder. As it was mentioned in Section 1 the representation of the fields in the form (2), (3) and (4) can be considered as continuous modes of a circular dielectric rod [2]. These modes have a continuous eigenvalue spectrum and the eigenvalues of the propagating continuous modes correspond to the real values of the angle of incidence $\theta$ [2]. In such a way, it can be assumed that the resonances with Fano like profiles occur when there is the interference between the modes of continuous spectrum and some discreet surface waves of the single cylinder in the vicinity of transverse resonances of these waves which are determined by an equation $J_n(x)=0$. Such interaction can be described by the terms of $A_0 = \frac{1}{x}\frac{J_0'(x)}{J_0(x)}$ and $B_0 = \frac{1}{y}\frac{J_0'(y)}{J_0(y)}$, $C_0 = \frac{1}{y}\frac{H_n^{(2)'}(y)}{H_0^{(2)}(y)}$, correspondingly. It is worth mentioning here that the amplitude of $a_0^s$ in (14) is determined not only by diffraction order of $J_0(x)$ but also by $J_1(x)$ due to the properties of cylindrical Bessel functions [4].

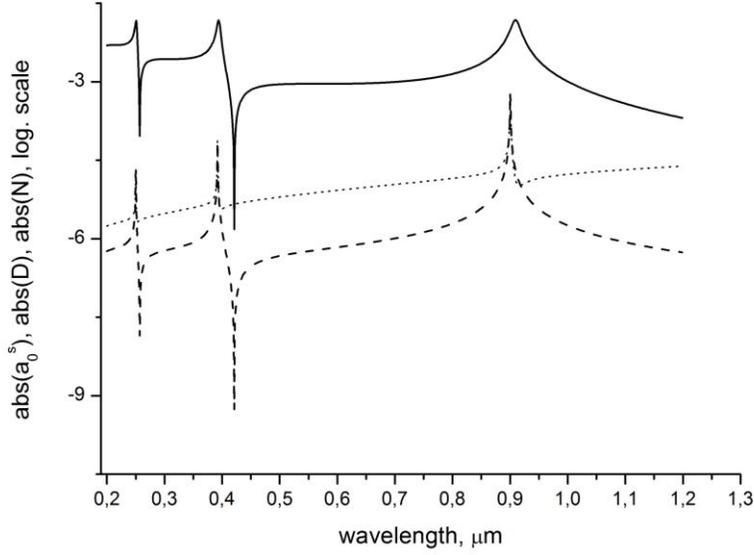

Fig. 2. The spectral dependencies of $abs(a_0^s)$ (solid), an absolute value of the denominator (*D*) in (14) (dotted) and the numerator (*N*) in (14) (dashed).

The spectral dependencies of $A_0(\lambda)$, $B_0(\lambda)$ and $C_0(\lambda)$ are shown in Fig. 3.

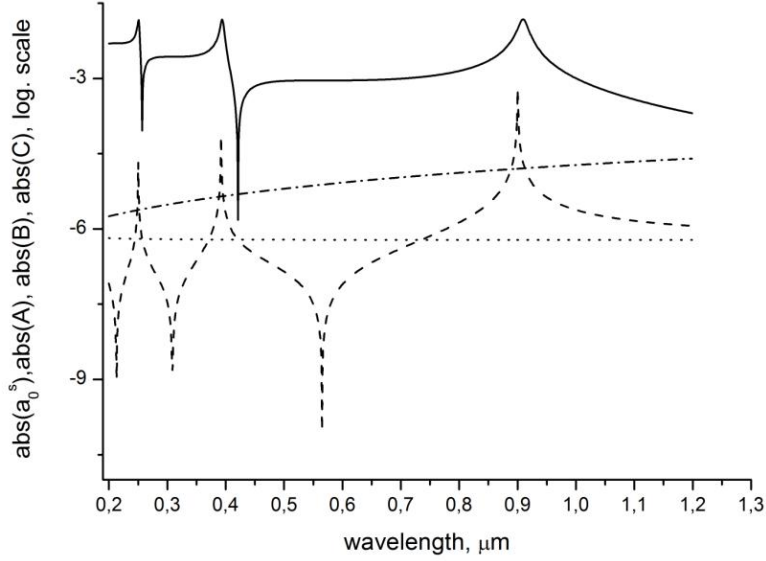

Fig. 3. The spectral dependencies of $abs(a_0^s)$ (solid), $abs(A_0(\lambda))$ (dashed), $abs(B_0(\lambda))$ (dotted) and $abs(C_0(\lambda))$ (dash - dotted).

It is seen from Fig. 3 that a curve of $abs(A_0(\lambda))$ intersects a curve of $abs(B_0(\lambda))$ at wavelengths corresponding to minimums of the Fano like profiles of $abs(a_0^s)$ ($\lambda = 0.42$ μm and $\lambda = 0.256$ μm). At the same time, all maximums of the Fano like and Lorentzian resonances of $abs(a_0^s)$ correspond to the maximums of $abs(A_0(\lambda))$ (Fig. 3). As it was mentioned above, these maximums are determined by the equation $J_0(x)=0$ for the transverse resonances in the cylinder or, in other words, by the poles of expression $A_0(\lambda)$. The curve of $abs(C_0(\lambda))$ intersect the curve of $abs(A(\lambda))$ at the wavelengths which don't correspond to any minimum or maximum of the Fano like resonances of ($abs(a_0^s)$). In such a way, it is the numerator which plays a decisive role in forming the Fano like profiles of the resonances of ($abs(a_0^s)$). It is also seen that the Fano like resonances occur only in proximity to the poles of $A(\lambda)$ ($J_0(x)=0$) and don't occur in proximity of the zeros which determined by an equation of $J_1(x)=0$.

To understand the mechanism of the Fano like resonances formation it is necessary to draw an analogy with plane wave scattering from 1D diffraction grating [10]. As it was pointed out in Section 2 the Wood anomalies connected with Rayleigh wavelengths occur when zero of the numerator of equation (13) is in close proximity to the pole of this equation. In our case, the numerator of (14) can be equal to zero when the spectral dependencies of $A_0 = \frac{1}{x}\frac{J_0'(x)}{J_0(x)} = -\frac{1}{x}\frac{J_1(x)}{J_0(x)}$ and $B_0 = \frac{1}{y}\frac{J_0'(y)}{J_0(y)} = -\frac{1}{x}\frac{J_1(y)}{J_0(y)}$ intersect at some value of wavelength and have the same sign. The arguments $x$ and $y$ are real due to real value of the angle of incidence of the plane wave as it was mentioned above. In Fig. 4 the spectral dependencies for the terms of $J_0(x)$, $J_1(x)$ and $J_0(y)$, $J_1(y)$ are shown.

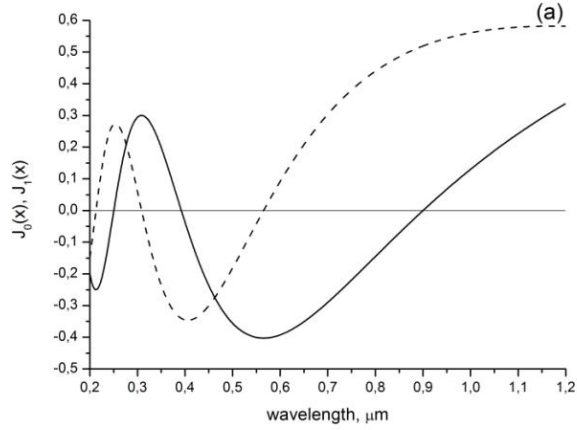

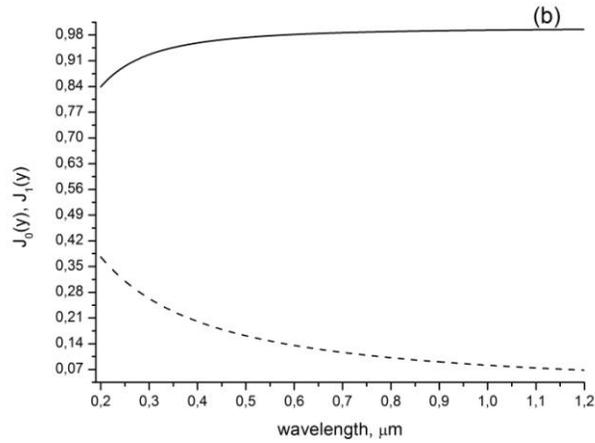

Fig.4. (a) the spectral dependencies of the terms of $J_0(x)$ (solid) and $J_1(x)$ (dashed); (b) the same dependencies for $J_0(y)$ (solid) and $J_1(y)$ (dashed).

It is seen that in the spectral region $\lambda > 0.9$ μm (the first pole of $A(\lambda)$) $J_0(x)$ and $J_1(x)$ have the same sign but $abs(A_0(\lambda)) > abs(B_0(\lambda))$ (Fig. 3). The functions of $J_0(y)$ and $J_1(y)$ have the same sign in all considered spectral region (Fig. 4(b)). It means that the first CSR at $\lambda = 0.9$ μm is determined only by the first pole of $A_0(\lambda)$ and have Lorentzian profile. At the same time, the spectral dependence of $C_0(\lambda)$ have a cross point with $A_0(\lambda)$ in this spectral range (Fig. 3). It means that a denominator approaches a minimum at this point but not an exact zero (Fig. 2) because Hankel function $H_0^{(2)}(y)$ has an imaginary part connected with function $Y_0(y)$. Due to this fact one can observe only a faint minimum of the Fano like profile for the denominator in this spectral range (Fig. 2). After the first pole of $A_0(\lambda)$ up to the first zero of $J_1(x)$ both functions of $J_0(x)$ and $J_1(x)$ have different signs and the numerator cannot be equal to zero at cross point $\lambda = 0.75$ μm (Fig. 3). Due to this fact the numerator has no any minimum at this point and spectral dependence of $abs(a_0^s)$ is smooth (Fig. 2).

The most interesting behavior of the numerator is observed when $B_0(\lambda)$ intersects $A_0(\lambda)$ in the vicinity of the second pole of $A_0(\lambda)$ at $\lambda = 0.4$ μm (Fig. 3). In this case, at $\lambda > 0.4$ μm there is a cross point between $A_0(\lambda)$ and $B_0(\lambda)$ (Fig. 3) and the signs of $J_0(x)$ and $J_1(x)$ are the same (Fig. 4(a)). It means that the numerator in (14) has a zero at this cross point and spectral dependence of $abs(a_0^s)$ must have a minimum (Fig. 2). This minimum is in close proximity to the second pole of $A_0(\lambda)$ and to the maximum of expression (14) (Fig. 3), correspondingly, as in the case of 1D diffraction grating (13). Due to this fact the resonance in this spectral region has Fano like profile (Fig. 3). Other CSRs of $abs(a_0^s)$ can be explained in the same way.

More complicated structure of the resonances is observed for the spectral dependencies of higher diffraction orders of the scattered magnetic field, for example, for $abs(a_1^s)$ (Fig. 1(a)). Resonances of such type (Fig. 1(a) (dashed)) were also obtained for 1D diffraction grating in [10]. To understand an occurrence of CSRs with such profiles (Fig. 1(a)) it is necessary to consider again the spectral dependencies of the numerator and denominator of $abs(a_1^s)$ separately. The spectral dependencies of the numerator and denominator are shown in Fig. 5.

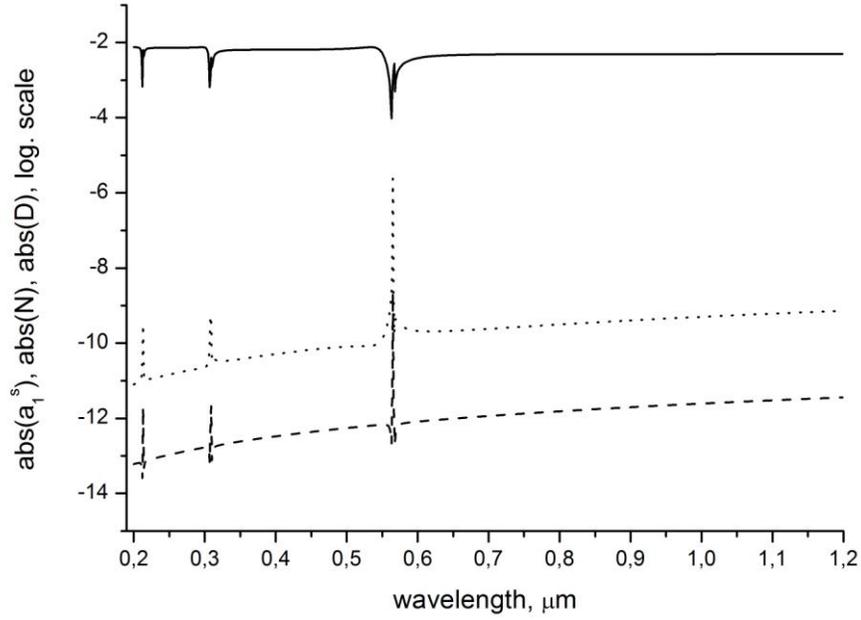

Fig. 5. The spectral dependencies of $abs(a_1^s)$ (solid), an absolute value of the denominator ($D$) in (9(a)) (dotted) and the numerator ($N$) in (9(a)) (dashed).

As one can see, all minimums of the spectral dependence of $abs(a_1^s)$ correspond to the maximums of the denominator which, in turn, correspond to poles of the term $\frac{1}{x}\frac{J_n'(x)}{J_n(x)}$ (9(a)) in the case of $n = 1$ (Fig. 5 (dotted)). All additional peaks in the narrow spectral regions corresponding to the minimums of $abs(a_1^s)$ occur due to maximums of the numerator (9(a)) in the case of $n = 1$ (Fig. 5 (dashed)). To clarify a process of forming this type of resonances one considers the spectral dependence of the numerator in 9(a) for $n = 1$. Its spectral behavior is determined by the resonances of both terms $A_1 = \frac{1}{y}\frac{J_1'(y)}{J_1(y)} - \frac{1}{x}\frac{J_1'(x)}{J_1(x)}$ and $B_1 = \frac{n_1^2}{x}\frac{J_1'(x)}{J_1(x)} - \frac{n_2^2}{y}\frac{H_1^{(2)'}(y)}{H_1^{(2)}(y)}$ in (9(a)).

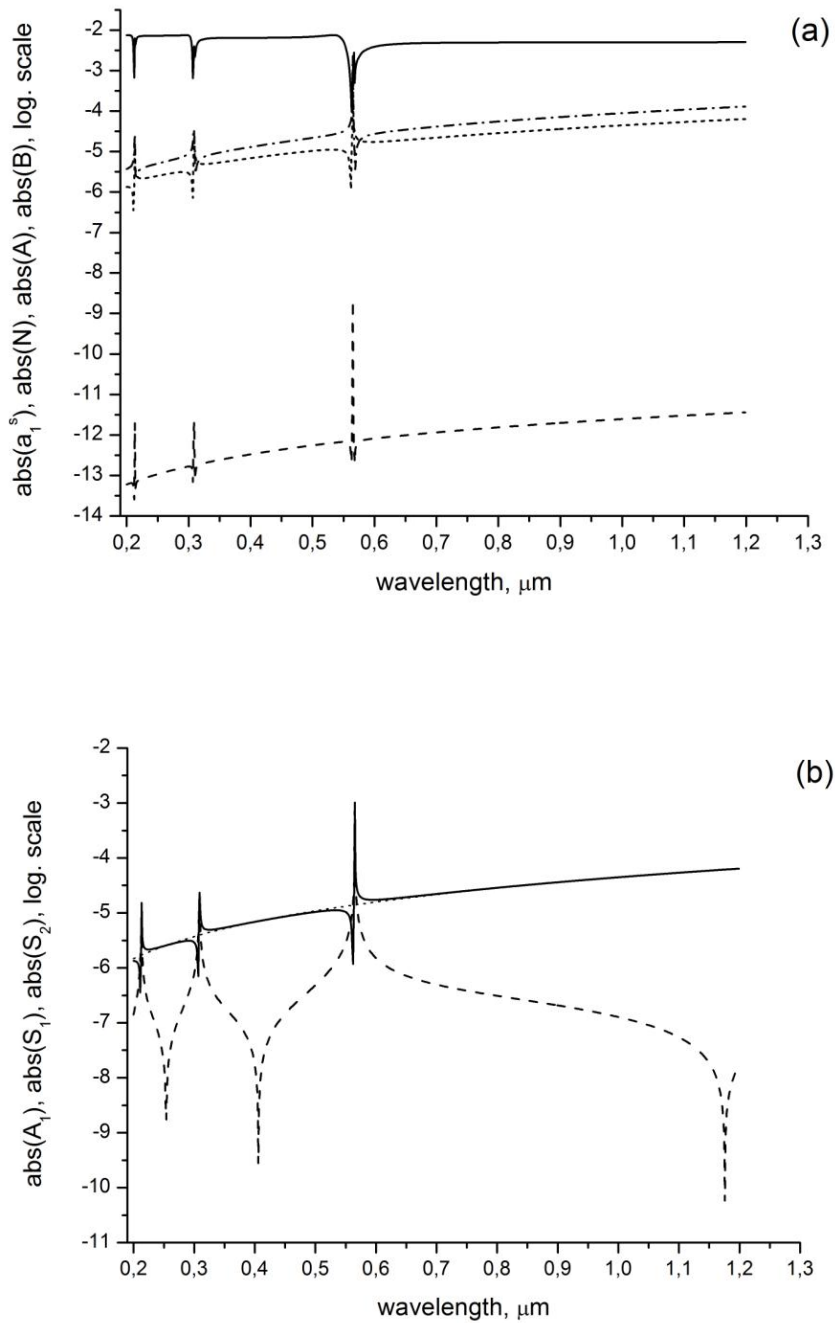

Fig. 6. (a) the spectral dependencies of $abs(a_1^s)$ (solid), an absolute value of the numerator ($N$) (dashed) and absolute values of $A_1$ (dotted) and $B_1$ (dashed - dotted); (b) the spectral dependencies of $abs(A_1)$ (solid), $abs(S_1)$ (dashed) and $abs(S_2)$ (dotted).

As one can see from Fig. 6(a), the resonances of the numerator have a complicated profile due to a superposition of two resonance terms of $A_1$ and $B_1$ which are Fano like resonances. These Fano like resonances are formed by interference between the summands in terms of $A_1$ and $B_1$. They have maximums at the poles of expressions $S_1 = \frac{1}{x}\frac{J_1'(x)}{J_1(x)}$ or $\frac{n_1^2}{x}\frac{J_1'(x)}{J_1(x)}$ and minimums at the intersection points between the spectral dependencies of $S_1$ and the spectral dependencies of $S_2 = \frac{1}{y}\frac{J_1'(y)}{J_1(y)}$ or $\frac{n_2^2}{y}\frac{H_1^{(2)'}(y)}{H_1^{(2)}(y)}$ at which the signs of $S_1$ and $S_2$ are different. The same situation was observed in the case of resonances occurred under excitation of CSW with $n = 0$ considered above (Fig. 3). To illustrate this fact the spectral dependencies of $S_1 = abs\left(\frac{1}{x}\frac{J_1'(x)}{J_1(x)}\right)$, $S_2 = abs\left(\frac{1}{y}\frac{J_1'(y)}{J_1(y)}\right)$ and $abs(A_1)$ are shown in Fig. 6(b).

Other CSRs (Fig. 1(a, b)) occurred under an excitation of CSWs on the cylinder surface including the high diffraction orders ($n > 1$) (Fig. 1) are also described in the terms of Lorentzian and Fano like resonaces and their origin are explained in the same way. The same type of CSRs and surface long range guided waves are excited at oblique incidence of the plane wave on any single dielectric cylinder with other values of $n_1$, $n_2$ and $a$. In the last section, we will analyze the main differences between the spectral dependencies of the amplitudes $abs(a_n^s(\lambda))$ for different values of ratios $a/\lambda$ and show how the quality factor of CSRs depends on the angle of incidence of the plane wave.

## The conditions of an effective excitation of CSWs and CSRs under plane wave scattering from a single dielectric cylinder

In this section one considers the conditions which are necessary for an effective excitation of different orders of CSRs and CSWs. We will also compare our results with the ones obtained in [5]. The authors of [5] obtained the grazing solution for the long ranged guided surface waves and pointed out that only CSWs with $n = 0, 1$ have a great impact on the resonance effects at the near – zero grazing incidence under the plane wave scattering from a single nanorod ($a \ll \lambda$). Historically, CSW with $n = 0$ was only considered in the radio frequency region due to high losses of CSWs of higher diffraction orders. In [5] it was shown that CWS with $n = 1$ begin to play a significant role in the optical regime at the near grazing incidence. As we have shown above (Fig. 1), CWSs of higher orders ($n > 1$) begin to play a significant role under increasing in the value of cylinder radius ($a \approx \lambda$). For example, CSWs with $n = 2, 3$ give strong resonance peaks in the short wavelength region in Fig. 1 which are comparable with magnitudes of the resonance peaks corresponding to CWSs with $n = 0,1$. In such a way, there is a direct connection between the value of ratio $a/\lambda$ and the number of excited CSWs at the near grazing incidence.

One of the main factors which play a decisive role in the excitation of the high quality CSRs is an angle of incidence of the plane wave. It is known [5] that the scattered waves and excited long range guided surface waves are interconnected at grazing incidence. In [5] a strong dependence of quality factor of CSRs on the angle of incidence was observed. In other words, the angle of incidence determines an effectiveness of the coupling between the coherently scattered continuous modes of the dielectric cylinder and its discreet propagating modes. In our case, the spectral dependencies of $abs(a_1^s)$ at four values of the angle of incidence are shown in Fig. 7.

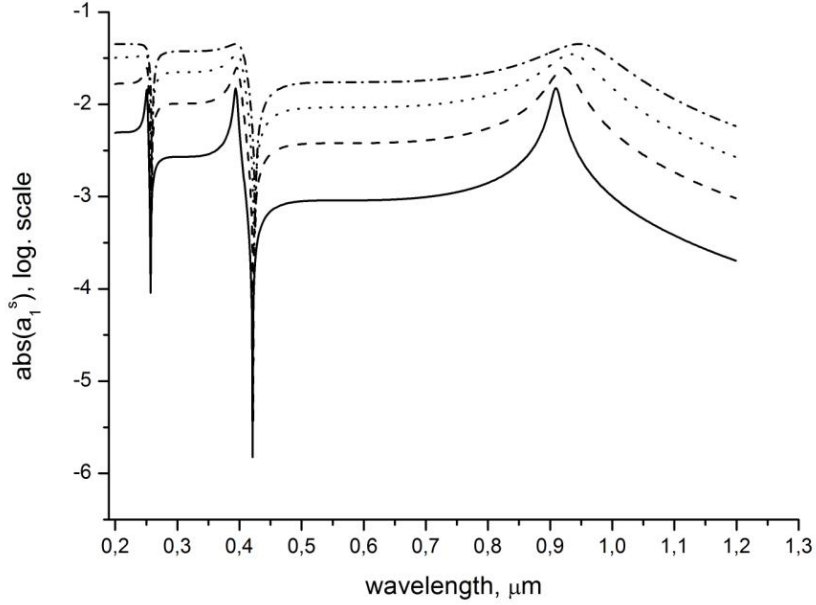

Fig. 7. The spectral dependencies of $abs(a_1^s)$ for four values of the angle of incidence of the plane wave at $\theta = 0.86^0$ (solid), $1.43^0$ (dashed), $2^0$ (dotted) and $3.6^0$ (dashed - dotted).

As one can see from Fig. 7, the quality factors of CSRs become lower with increasing in the value of the angle of incidence. It means that the efficiency of the coupling between the continuous modes of the dielectric cylinder and CSWs decreases and long ranged CSWs cannot be excited at greater values of the angle of incidence. It is not possible also to excite CSRs with high quality factor because the regime of CSWs excitation is destroyed. This result is also confirmed in [5] it was shown that effective waveguide regime for CSWs with high quality factor can exist only for the oblique incidence of the plane wave when $\theta \approx 0.01^0$. This fact demonstrates that the excitation of CSWs on the nanorod surface connected with lower values of the angle of incidence with respect to the case considered above $\theta \approx 1^0$ (Fig. 1). It can be connected with a fact that the radiation propagating along the nanorod is mostly concentrated in the scattered waves outside the nanorod and an effective excitation of CSWs and CSRs with high quality factor is possible only at such small values of the angle of incidence when the scattered continuum modes and the propagating modes are phase matched in $z$ - direction. In our case (Fig.1) $a \approx \lambda$, the interconnection between the scattered continuum modes and the propagating modes of the dielectric cylinder is stronger and the phase matching condition in $z$ - direction is weakened and the excitation of CSRs with high quality factor can be obtained at greater values of the angle of incidence.

## Conclusions

In conclusion, we have investigated a process of excitation of CSWs and CSRs under plane wave scattering from a single dielectric cylinder at oblique incidence of light. It was shown that the resonances (CSRs) occurred in the spectral dependencies of the absolute values of amplitudes of the scattered fields are analogous to the resonances occurred under plane wave scattering from 1D metallic diffraction grating (Wood's anomalies). The anomalies with Fano like profiles occur under an interaction of the continuous modes of cylinder with the discrete surface waves in the vicinity of the wavelength at which the condition of transverse resonance is satisfied ($J_n(x)=0$). In the case of 1D diffraction grating their location in the spectrum corresponds to Rayleigh wavelengths. The resonances with Lorentzian profile occur when such interaction is absent or can be neglected and only the condition of the transverse resonance is satisfied. In the case of 1D diffraction grating such resonances correspond to the type of the resonance anomalies which occur under excitation of the guided complex wave supportable by the grating. Moreover, there exist more complicated resonance profiles which occur for CSW with diffraction order $n=1$ and analogous resonances also exist in the case of 1D diffraction grating. All these facts lead to the conclusion that the curved cylindrical surface can be considered as an azimuthal quasi diffraction grating which transfers a momentum to the incident wave along the azimuth as 1D metallic diffraction grating transfer a momentum to the incident plane wave along the wave propagation direction ($z$ - axis). The number of excited CSWs and quality factor of CSRs (or an efficiency of CSWs excitation) depends on the ratio $a/\lambda$ and the angle of incidence of the plane wave. These conclusions can help to understand more deeply the process of light interaction with the cladding in microstructured optical fibers, for example, photonic band gap fibers and microstructured optical fibers with negative curvature of the core boundary.